\documentclass[sort&compress]{elsarticle}

\biboptions{sort&compress}
\usepackage{amsmath}
\usepackage{bm}
\usepackage{url}
\usepackage{epsfig}
\usepackage{color, colortbl}
\usepackage{arydshln}
\usepackage{rotating}
\definecolor{LightCyan}{rgb}{0.88,1,1}

\begin{document}

\newcommand{\rtotresult}{${\cal{R}}_0^{\rm tot}=11.0$ with 95\% CI $[9.4,15.6]$} 
\newcommand{\rtotresultb}{${\cal{R}}_0^{\rm tot}=11.0$ $[9.4,15.6]$}

\newcommand{\renvresult}{${\cal{R}}_0^{\rm environ}=0.85$ with 95\% CI $[0.18,2.04]$} 
\newcommand{\renvresultb}{${\cal{R}}_0^{\rm environ}=0.85$ $[0.18,2.04]$} 
\newcommand{\renvresultc}{$[0.18,2.04]$}

\newcommand{\rdirectresult}{${\cal{R}}_0^{\rm direct}=10.4$ with 95\% CI $[8.8,14.9]$} 
\newcommand{\rdirectresultb}{${\cal{R}}_0^{\rm direct}=10.4$ $[8.8,14.9]$}

\begin{frontmatter}
\title{
Quantifying the relative effects of environmental and direct transmission of
norovirus 
}

\author[st]{S.Towers\corref{cor1}}
\ead{smtowers@asu.edu}
\author[st]{J.Chen}
\author[st]{C.Cruz}
\author[st]{S.Madler}
\author[st]{J.Melendez}
\author[st]{J.Rodriguez}
\author[st]{A.Salinas}
\author[st]{F.Yu}
\author[st]{Y.Kang}

\cortext[cor1]{Corresponding Author}

\address[st]{Simon A. Levin Mathematical, Computational and Modeling Sciences Center,Arizona State University, Tempe, AZ, USA}

\begin{abstract}
{\bf Background:}
Norovirus is a common
cause of outbreaks of acute gastroenteritis
in health- and child-care settings,
with serial outbreaks also frequently observed aboard cruise ships.
The relative contributions of environmental and direct person-to-person transmission of norovirus has hitherto not been quantified. \\
{\bf Objective:}
We employ a novel mathematical model of norovirus transmission,
and fit the model to daily incidence data from a major norovirus
outbreak on a cruise ship, and examine the relative efficacy
of potential control strategies aimed at reducing environmental and/or direct transmission. \\
{\bf Results:}
The reproduction number 
for environmental and direct transmission combined is \rtotresultb,
and of environmental transmission alone is \renvresultb.
Direct transmission is overwhelmingly due to passenger-to-passenger contacts, but
crew can act as a reservoir of infection from cruise-to-cruise. \\
{\bf Implications:}
This is the first quantification of the relative roles of 
environmental and direct transmission of norovirus.
While environmental transmission has the potential to maintain a sustained series of outbreaks
aboard a cruise ship in the absence of strict sanitation practices, direct transmission dominates.
Quarantine of ill passengers and cleaning are likely to have little impact on final outbreak size, but intensive promotion of good hand washing practices can prevent outbreaks.

\end{abstract}
\end{frontmatter}


\section{Introduction}

Noroviruses are a group of non-enveloped, single-stranded RNA viruses~\cite{hall2011updated,weinstein2008gastrointestinal}, and are a highly infectious causal agent of sporadic and epidemic gastroenteritis~\cite{patel2009noroviruses, glass2009norovirus}; they are the most common cause of gastroenteritis, food-borne disease, and community acquired diarrheal disease across all ages in the United States~\cite{hall2011updated,robilotti2015norovirus,kaplan1982frequency}. 
Noroviruses are spread primarily through faecal-oral transmission, exposure to vomit, consumption of food or drink that has been contaminated, contact with contaminated surfaces, or direct contact with those infected~\cite{lopman2012environmental,kampmeier2016norovirus}.
Each year in the United States, noroviruses cause, on average, 19-21 million cases of acute gastroenteritis \cite{hall2013norovirus}, leading to 1.7-1.9 million outpatient visits and 400,000 emergency department visits, 56,000-71,000 hospitalisations, and 570-800 deaths, mostly among young children and the elderly~\cite{hall2013norovirus}.
The frequent occurrence of outbreaks has resulted in the substantial economic burden of \$500 million for norovirus associated hospitalisations in the United States~\cite{morris2011ranking}. 

One of the key challenges that noroviruses pose is short lived immunity with limited cross-protection between strains, enabling multiple potential infections with the viruses through the lifetime of the host~\cite{patel2009noroviruses, glass2009norovirus,simmons2013duration}.
Additionally, the environmental durability of norovirus leads to persistence of the pathogen in clinical settings, and other closed or semi-closed environments such as daycare centres, schools, and cruise ships, thus complicating complete disinfection and allowing for recurrent outbreaks~\cite{robilotti2015norovirus,cheesbrough2000widespread,wikswo2011disease}.
There are currently no vaccines or specific treatments (other than palliative treatment) 
for noroviruses, leading to sanitation, personal hygiene practices, and quarantine or isolation as being the only potential
means of control of the spread of the disease~\cite{koo2010noroviruses}.

Cruise ships, in particular, have seen an increase in norovirus outbreaks in recent years, coinciding with the increased popularity of cruise
vacations~\cite{hadjichristodoulou2011surveillance,bert2014norovirus,wikswo2011disease,lawrence2004outbreaks}. 
With around 25 million passengers annually across the world and growing, almost one million full-time equivalent jobs with \$38 billion in wage and salaries, and over \$100 billion economic impact worldwide, the cruise industry has an influence on the lives of people in terms of recreation and employment~\cite{CLIA}.  
Unfortunately, the environment on cruise ships has the ingredients of a `perfect storm' for outbreaks, with common food and drink, shared spaces for most activities, and a semi-closed environment~\cite{bert2014norovirus,robilotti2015norovirus},  
and the economic disincentive for passengers and crew to report illness may also play a role in complicating control of outbreaks~\cite{robilotti2015norovirus}.
Even with good performance on environmental health sanitation inspections, it has been noted that
outbreaks of gastroenteritis per 1000 cruises have been increasing in time~\cite{cramer2006epidemiology}, indicating that sanitation is not the only factor that must be considered in outbreak control. 

Statistical analyses based on survey questionnaires have examined the relationships between certain behaviours and norovirus outbreaks. For example, a systematic review of the literature associated with 127 past norovirus outbreaks on cruise ships, showed that preventative measures, such as food hygiene or the promotion of preventative procedures, could potentially affect outcomes~\cite{bert2014norovirus}.  Another survey found that passengers with a lackadaisical attitude toward hand hygiene were more likely to be infected with norovirus during an outbreak~\cite{wikswo2011disease}.

Beyond statistical analyses, however, mathematical models that describe the dynamics of the spread of a pathogen can yield insights into the relative efficacy of potential control measures~\cite{hethcote2000mathematics}.
There have been several previous works that explore the dynamics of the spread of norovirus:
\begin{itemize}
\item Vanderpas {\it et al.} (2008) constructed a simple deterministic compartment model to analyse a norovirus outbreak in long-term care facilities in a closed population~\cite{vanderpas2009mathematical}, which highlighted the need for reinforced infection control measures. 
\item Simmons {\it et al.} (2013) constructed an age-structured deterministic transmission model to examine the duration
of immunity after infection, and the age-dependence of transmission~\cite{simmons2013duration}.
\item Zelner {\it et al.} (2013) examined a compartmental model that incorporated temporal changes in infectiousness, to examine the implications for control of transmission in a household setting.
\item Bartsch {\it et al.} (2014) constructed an agent-based model to simulate the spread of norovirus within and between 29 acute care hospitals and 5 long-term care facilities~\cite{bartsch2014spread}, and concluded that control was better achieved when hospitals acted cooperatively to track outbreaks. 
\item Lopman {\it et al.} (2014) developed a dynamic transmission model of norovirus infection, disease, and immunity to gain an understanding of the apparent high prevalence of asymptomatic infection~\cite{lopman2014epidemiologic}.
\item Assab and Temmime (2016) used a stochastic model to examine the effects of hand washing and isolation on the spread of norovirus within nursing homes. 
\item Lee {\it et al.} (2016) developed simulation models to determine the
potential cost-savings from the hospital perspective of 
implementing various norovirus outbreak control interventions~\cite{lee2011economic}. 
\end{itemize}

However, until now, no norovirus model has examined the relative contributions of environmental and person-to-person transmission to the overall transmissibility of the virus within a population.  In addition, no model has incorporated the unique temporal dynamics of outbreaks on cruise
ships, which can exhibit multiple waves from cruise-to-cruise as the passenger population refreshes with each new cruise, while the
crew population remains the same, potentially acting as a reservoir of infection.
Incorporation of these dynamics into a mathematical model of norovirus transmission can aid in assessing the relative efficacy of
control strategies aimed at sanitation, hygiene, isolation, and/or quarantine of sick patients.
The main objectives of our analysis are three-fold:
\begin{enumerate}
        \item To explore the dynamical effects of both environmental and direct transmission
of norovirus on cruise ships by constructing a realistic mathematical model.
        \item Optimise the model parameters to outbreak data.
        \item Use the model to provide insights to help inform effective control strategies in a cruise ship setting.
\end{enumerate}

In the following sections, we will describe our data and mathematical model, followed by a presentation of results and discussion.

\section{Methods and Materials}
\subsection{Data}

\begin{figure}[h]
   \begin{center}
    \mbox{\put(-190,0){ \epsfxsize=13cm
           \epsffile{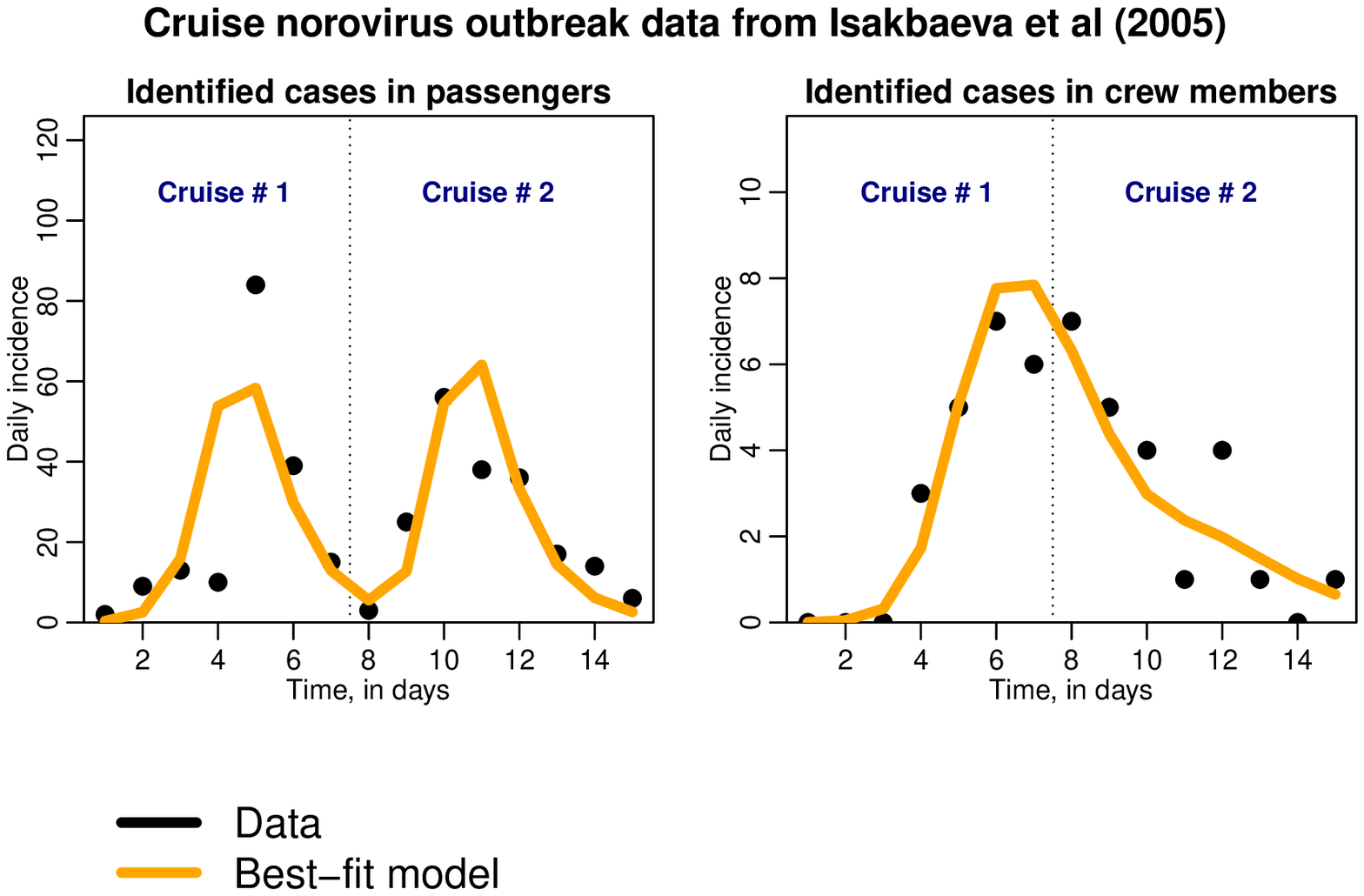}
     }}
     \vspace*{-0.0cm}
  \caption{
      \label{fig:results}
  Time series of reported acute gastroenteritis cases, by date of symptom onset, for passengers
 and crew aboard two consecutive cruises aboard a cruise ship during late 2002, from Reference~\cite{isakbaeva2005norovirus}.
Overlaid is the best-fit model of Equations~\ref{eqn:model} that includes both direct and environmental
transmission.
   }
\end{center}
\end{figure}

Data for these studies were the time series of the daily number of identified cases of acute gastroenteritis
(AG)
among 2,300 to 2,400 passengers and 999 crew members during a six week period
aboard a cruise ship in late 2002~\cite{isakbaeva2005norovirus}.
Weekly cruises took place, with new passengers at each cruise, but the crew members remaining the same from cruise-to-cruise.
The first two cruises recorded the largest number of AG cases, which laboratory testing of stool samples revealed to
be primarily caused by a single strain of norovirus.  The ship was taken out of service for a week after the first two
cruises for thorough sanitation, but AG cases continued (although at a decreased level) for several subsequent cruises.
Laboratory analysis of stool samples from identified AG cases
in the later cruises revealed that six different norovirus strains were
involved.
We confine this analysis to examination of the time series data from the first two cruises, which involved the
single strain of norovirus, to avoid the complications of taking into account cross-immunity between strains.
The time series data for passengers and crew members are shown in Figure~\ref{fig:results}.


From a survey study of the identified cases, the investigators of the outbreak concluded that the 
initial index case(s) may have been due to potential food-contamination, but that
secondary cases were likely due to person-to-person infection (with the crew potentially acting as reservoir of
infection from cruise-to-cruise), and potential environmental fomite transmission~\cite{isakbaeva2005norovirus}.

\subsection{Mathematical Model}

In this study, we employ a deterministic Susceptible, Exposed, Infected, and Recovered (SEIR) type mathematical
model to simulate the transmission dynamics of norovirus on a cruise ship. 
Similar models have been used in past studies of norovirus transmission~\cite{vanderpas2009mathematical,lopman2014epidemiologic},
and have also been used to examine a wide
array of diseases, including influenza, measles, and Ebola, to name but a few~\cite{grassly2008mathematical}.
We modify the model to include environmental contamination, via an additional model 
compartment, $W$, similar to some cholera models~\cite{tien2010multiple,chao2013modeling}.

Additionally,
we incorporate the potentially differing direct transmission dynamics within, and between, the 
two primary sub-populations on a cruise ship; the crew and passengers.  
In the model, the crew members contact (and can be infected upon exposure) infectious crew members or passengers, 
whereupon they move to the exposed, but not yet infectious,
compartment.  After a period of time, $1/\kappa$, they proceed to the infectious compartment.  They recover with
immunity after $1/\gamma$ days. Because a cruise season is much shorter than the typical duration of immunity 
upon recovery from norovirus, we ignore waning immunity in the model.  We also ignore births and deaths in 
the model.
Similar to the crew, passengers also contact crew and other passengers, with potentially differing rates compared
to the contacts the crew make, and can be infected by infectious individuals in both groups.

The inclusion of both environmental and direct transmission is novel for modelling studies of the transmission of norovirus in any setting, 
and the inclusion of
the dynamics of interaction between crew and passengers is novel for the study of norovirus transmission aboard cruise ships.

The differential equations describing these dynamics are:
\begin{eqnarray}
{{dS_i}\over{dt}} &=& -\eta_W W S_i  -S_i \sum_j B_{ij} I_j/N_j \nonumber \\
{{dE_i}\over{dt}} &=& +\eta_W W S_i  +S_i \sum_j B_{ij} I_j/N_j - \kappa E_i \nonumber \\
{{dI_i}\over{dt}} &=& +\kappa E_i - \gamma I_i \nonumber \\
{{dR_i}\over{dt}} &=& +\gamma I_i \nonumber \\
{{dW}\over{dt}} &=& +\alpha\sum I_i - \xi W,
\label{eqn:model_tota}
\end{eqnarray}
where the sub-population indices, $i$ and $j$, refer to ``passengers'' when $i,j=1$, and ``crew'' when $i,j=2$.
The parameter $B_{ij}$ is the contact rate, sufficient to transmit infection, between
individuals in sub-population $i$, and those in sub-population $j$. 
The parameter
$\eta_W$ is the rate at which the population contacts the environment, 
and $\alpha$ and $\xi$ are the excretion and decay rates of the pathogen into, and out of, the environment, 
respectively.
We also have the population size
$N=N_1+N_2=S_1+E_1+I_1+R_1+S_2+E_2 + I_2 + R_2$.

Because the population is closed, the amount of time passengers spend with crew must equal the amount of time crew spends with passengers,
thus under the assumption that the probability of transmission upon contact is the same for
both groups,
the transmission matrix must satisfy reciprocity~\cite{wallinga2006using}, which means that 
${{N_{1}}} B_{12} = {{N_2}} B_{21}$.

Following Reference~\cite{tien2010multiple}, we re-scale the environmental compartment of Equations~\ref{eqn:model_tota}, 
such that $W_{\rm new}\rightarrow {{\xi}\over{\alpha N}} W_{\rm old}$.
This yields
\begin{eqnarray}
{{dS_i}\over{dt}} &=& -\beta_W W S_i  - S_i \sum_j B_{ij} I_j/N_j \nonumber \\
{{dE_i}\over{dt}} &=& +\beta_W W S_i  + S_i \sum_j B_{ij} I_j/N_j - \kappa E_i \nonumber \\
{{dI_i}\over{dt}} &=& +\kappa E_i - \gamma I_i \nonumber \\
{{dR_i}\over{dt}} &=& +\gamma I_i \nonumber \\
{{dW}\over{dt}} &=& +\xi\left(\sum I_i/N -  W\right),
\label{eqn:model}
\end{eqnarray}
with scaled environmental transmission rate $\beta_W = \eta_W N \alpha/\xi$.

\begin{figure}[h]
   \begin{center}
    \mbox{\put(-190,0){ \epsfxsize=13cm
           \epsffile{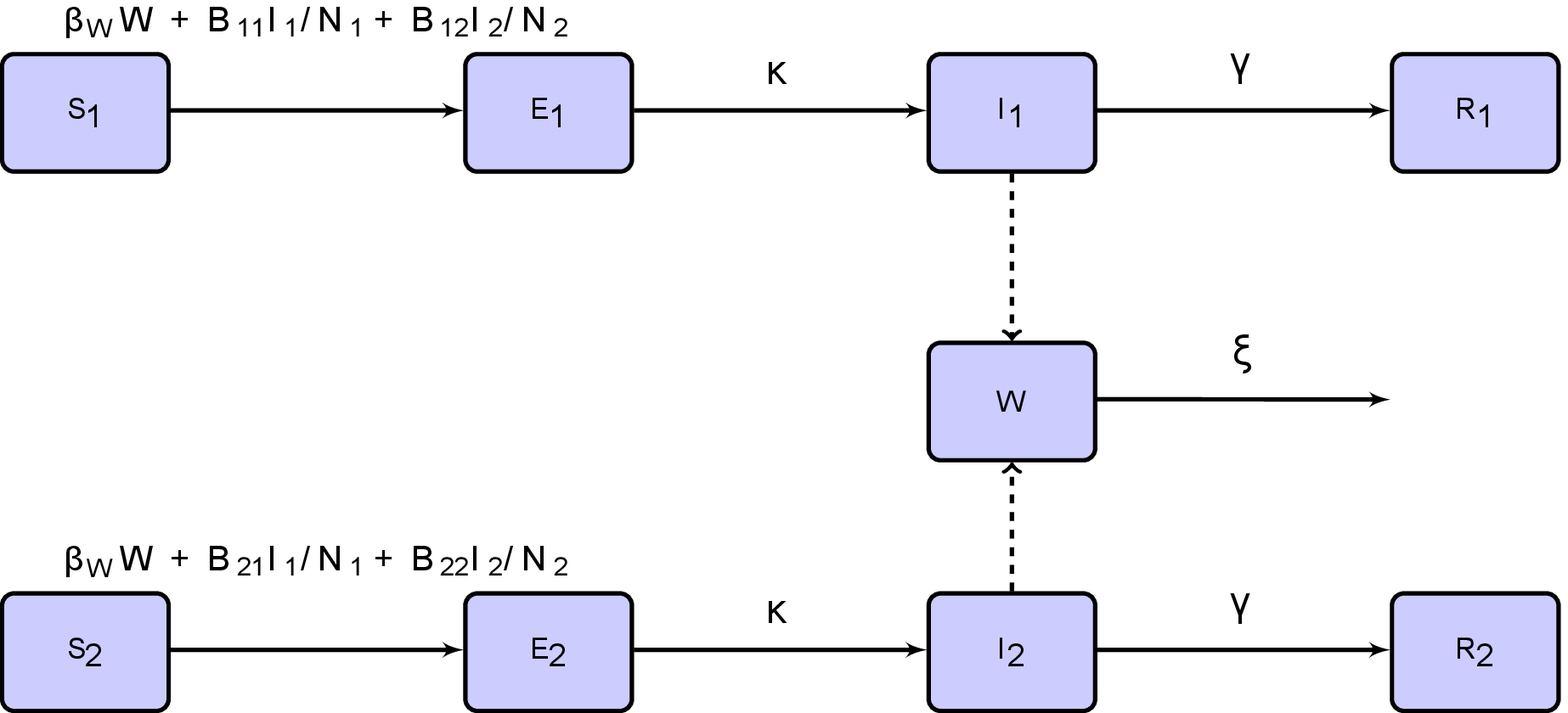}
     }}
     \vspace*{-0.0cm}
  \caption{
      \label{fig:compartment}
Compartmental flow diagram of the model described in Equations~\ref{eqn:model}.
In the diagram, subscripts 1 and 2 refer to the passenger and crew sub-populations, respectively.
   }
\end{center}
\end{figure}

The compartmental diagram for the model of Equations~\ref{eqn:model}
is shown in Figure~\ref{fig:compartment},
and a summary of the parameters of the model is given in Table~\ref{tab:parameters}.

The basic reproduction number of a disease is the average number of new infections produced by a single
infectious individual during the course of their infection, in a completely susceptible population~\cite{hethcote2000mathematics,diekmann2009construction}.
As described in Appendix A, 
using the next generation matrix approach~\cite{diekmann2009construction}, we find that
the reproduction number of the model of Equations~\ref{eqn:model} is 
\begin{eqnarray}
{\cal{R}}_0^{\rm tot}
            & = & {{B_{11}+B_{22}+\beta_W+A}\over{2\gamma}},
\label{eqn:r0}
\end{eqnarray}
where
\begin{eqnarray}
A  & = & \sqrt{(B_{11}-B_{22}+(f_1-f_2)\beta_W)^2 + 4*f_1*f_2(B_{12}/f_2+\beta_W)(B_{21}/f_1+\beta_W)},
\nonumber
\end{eqnarray}
and where $f_i$ is the fraction of the population in sub-group $i$.

In the absence of direct transmission (i.e. transmission occurs through the environment only), $B_{ij}=0$ for all $i$ and all $j$, and
Equation~\ref{eqn:r0} simplifies to
\begin{eqnarray}
{\cal{R}}_0^{\rm environ} & = & \beta_W/\gamma.
\label{eqn:R0_env}
\end{eqnarray}

\begin{sidewaystable}[t]
\begin{center}
\caption{
Parameters of the model of Equations~\ref{eqn:model}.
Quantities described below the dashed line are derivative of the parameters above the line.
\hskip 2in}\label{tab:parameters}
\begin{tabular}{llll}
  \hline
\\[-2mm]
Parameter    & Description  & Value & Reference\\[2mm] \hline
\rowcolor{LightCyan}
$N_1$ ($N_{\rm pass}$) & Number of passengers              & 2,318  & \cite{isakbaeva2005norovirus}\\
$N_2$ ($N_{\rm crew}$) & Number of crew                    & 999    & \cite{isakbaeva2005norovirus}\\
\rowcolor{LightCyan}
$B_{11}$    & Rate passengers contact passengers & TBD    & \\
$B_{12}$    & Rate passengers contact crew       & TBD    & \\
\rowcolor{LightCyan}
$B_{22}$    & Rate crew contact crew             & TBD    & \\
$B_{21}$    & Rate crew contact passengers       & $N_1 B_{11}/N_2$ & \cite{wallinga2006using}  \\
\rowcolor{LightCyan}
$\beta_W$    & Environmental transmission rate   & TBD    & \\
$1/\kappa$   & Incubation period                          & $[1.1,1.2]$ days & \cite{lee2013incubation}\\
\rowcolor{LightCyan}
$1/\gamma$   & Infectious period                          & $1.17$ $[1.00,1.88]$ days & \cite{zelner2010infections}\\
$\xi$        & Decay rate of norovirus in the environment      & $0.61$ $[0.38,0.84]$ days$^{-1}$ & \cite{mattison2007survival}\\[2mm]
\hdashline 
\\[-2mm]
\rowcolor{LightCyan}
${\cal{R}}_0^{\rm tot}$ & Basic reproduction number of direct and environmental transmission & & \\
${\cal{R}}_0^{\rm direct}$ & Basic reproduction number of direct transmission alone & & \\
\rowcolor{LightCyan}
${\cal{R}}_0^{\rm environ}$ & Basic reproduction number of environmental transmission alone &  & \\
\hline
\end{tabular}
\end{center}
\end{sidewaystable}


A cruise ship
has regular emigration and immigration of passengers (but not the crew).
We thus assume that the initial conditions at the beginning of the first cruise in the data are one infected passenger introduced to a population of 999 crew and 2,317 other passengers that are susceptible~\cite{isakbaeva2005norovirus}.  At the beginning of the subsequent cruise, the passengers are replaced with an entirely susceptible group of new passengers, while the crew population remains.  Because of the limited duration of the
cruises and the temporal variation in the refreshing of the passenger population with new groups of susceptible individuals, the model cannot be analytically analysed to estimate outbreak final size.  
Rather, we must rely on numerical simulations.

With our model, the efficacy of reducing environmental transmission through cleaning can be
examined (which results in an increase in the pathogen environmental decay rate, $\xi$, and a proportional decrease in $\beta_W$). 
We also examine interventions aimed at immediate and
complete quarantine of
some fraction, $f_{\rm quar}$, of infectious individuals; as described in Appendix A, and Reference~\cite{feng2007final}, this
results in a relative reduction in 
${\cal{R}}_0^{\rm tot}$ of $(1-f_{\rm quar})$.
We also examine interventions aimed at
reducing the overall contacts sufficient to transmit infection in the population, due, for example,
to personal hygiene practices designed to reduce the probability of transmission on contact either
directly with another person, or with a contaminated environmental surface (i.e. proportional reductions in both
$B_{ij}$ and $\beta_W$).

\subsection{Statistical methods}

In our analysis, we fit the transmission rate parameters $\beta_W$, $B_{11}$, $B_{12}$,
and $B_{22}$ of the model shown in Equations~\ref{eqn:model}
to the Isakbaeva {\it et al.} (2005) cruise outbreak data~\cite{isakbaeva2005norovirus}, 
using a Negative Binomial likelihood fit to account for over-dispersion in the 
data~\cite{lloyd2007maximum}.  

We also fit for the incubation and infectious periods, $1/\kappa$ and $1/\gamma$, respectively, with the
likelihood modified to include the Bayesian prior probability distributions for these values, as obtained
from previous studies:
\begin{itemize}
\item The incubation period of norovirus in a community setting has been estimated by some studies to be approximately 
2 days~\cite{weinstein2008gastrointestinal,rockx2002natural}, and by another meta-analysis 
study to have 95\% confidence interval $[1.1,1.2]$ days~\cite{lee2013incubation}.
\item There are few estimates of the infectious period of norovirus, although it is suspected it is longer than
the duration of symptoms~\cite{atmar2008norwalk,milbrath2013heterogeneity}.
An analysis of a norovirus community outbreak in Sweden estimated
that $1/\gamma$ is $1.17$ days with 95\% CI $[1.00,1.88]$~\cite{zelner2010infections}.
\end{itemize}

To incorporate this prior 
information into our likelihood fit, we assume that the prior probability distribution
for $1/\kappa$ is Normal, with mean $1.15$ days and standard deviation $0.1/1.96/2=0.03$ days.
We assume that the prior probability distribution for $1/\gamma$ is an
asymmetric Normal distribution, with mean $\mu=1.17$ days, and
standard deviation $\sigma=(1.17-1)/1.96=0.09$ days when $1/\gamma\le\mu$, and $\sigma=(1.88-1.17)/1.96=0.36$ days when
$1/\gamma>\mu$.
Given hypotheses for $1/\kappa$ and $1/\gamma$, 
the likelihood is then modified by multiplying the likelihood by the probabilities calculated from these
two distributions.

The rate of decay in viability of  norovirus on environmental surfaces is poorly known, 
largely because norovirus cannot be grown in cell culture~\cite{koo2010noroviruses,mattison2007survival}.
However, the feline calicivirus (FCV) has been used by several investigators as an acceptable surrogate
for norovirus in inactivation studies; Mattison {\it et al.} (2007) examined the
inactivation of FCV in food, and on metal at various temperatures~\cite{mattison2007survival}.
From the time series
data presented in Mattison {\it et al.} (2007) for FCV on metal surfaces at
room temperature, we estimate that the exponential rate of decline of the virus is $0.607\pm0.117$
days$^{-1}$.
In the fit of model Equations~\ref{eqn:model} to the cruise outbreak data, 
we thus also fit for the rate of decay of the virus in the environment,
$\xi$, with the likelihood modified with the Normal prior probability
distribution for $\xi$, with mean $0.607$ days$^{-1}$, and standard deviation $0.117$ days$^{-1}$.


\section{Results}

\subsection{Model optimisation}
In Table~\ref{tab:results} we show the 
results of the optimisation of the model parameters of Equations~\ref{eqn:model} to
the cruise outbreak data of Isakbaeva {\it et al.} (2005)~\cite{isakbaeva2005norovirus}.
In Figure~\ref{fig:results} we show the best-fit model of Equations~\ref{eqn:model} overlaid on the data.

\begin{table}[t]
\begin{center}
\caption{
Results of the optimisation of the model parameters of Equations~\ref{eqn:model} to
the cruise outbreak data of Isakbaeva {\it et al.} (2005)~\cite{isakbaeva2005norovirus}.
The table values below the dashed line are derivative of the model parameters above
the dashed line.
The best-fit model overlaid on the data is shown in Figure~\ref{fig:results}.
\hskip 2in}\label{tab:results}
\begin{tabular}{ll}
  \hline
\\[-2mm]
Parameter    & Best-fit and 95\% CI  \\[2mm] \hline
$B_{11}$       & $9.1$ $[7.4,10.9]$ days$^{-1}$\\
$B_{12}$          & $0.25$ $[0.03,0.6]$            \\
$B_{22}$          & $0.84$ $[0,2.39]$            \\
$1/\kappa$   & $1.15$ $[1.11,1.21]$ days       \\
$1/\gamma$   & $1.14$ $[1.01,1.74]$ days       \\
$\beta_W$    & $0.74$ $[0.13,1.78]$ days$^{-1}$\\
$\xi$        & $0.6$ $[0.39,0.82]$ days$^{-1}$\\[2mm] \hdashline 
\\[-2mm]
${\cal{R}}_0^{\rm tot}$         & $11$ $[9.4,15.6]$ \\
\\[-2mm]
${\cal{R}}_0^{\rm direct}$         & $10.4$ $[8.8,14.9]$ \\
${\cal{R}}_0^{\rm environ}$    & $0.85$ $[0.18,2.04]$ \\
$(B_{21}+B_{22})/(B_{11}+B_{12})$   & $0.15$ $[0.02,0.3]$ \\
$f{\rm confirmed}_{\rm passengers}$         & $0.0791$ $[0.0787,0.0797]$ \\
$f{\rm confirmed}_{\rm crew}$               & $0.046$ $[0.045,0.052]$ \\[2mm]
\hline
\end{tabular}
\end{center}
\end{table}

\subsection{Evaluation of potential control measures}

\begin{figure}[h]
   \begin{center}
    \mbox{\put(-190,0){ \epsfxsize=13cm
           \epsffile{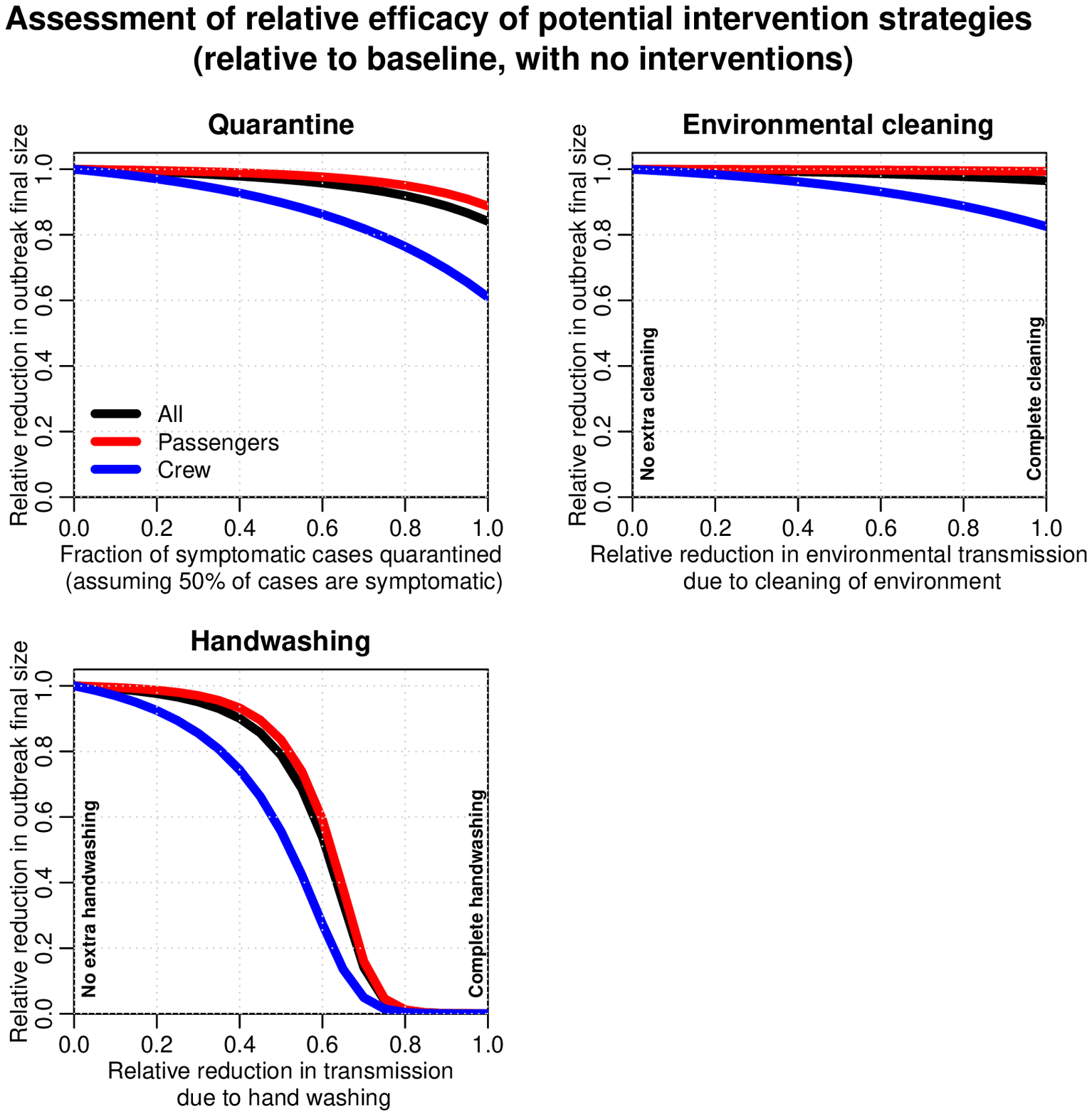}
     }}
     \vspace*{-0.0cm}
  \caption{
      \label{fig:interventions}
Effect of potential control strategies on outbreak final size, relative to the
observed baseline.  The black line represents the relative reduction for all individuals, while
the red and blue lines show the relative reductions for passengers and crew, respectively.
   }
\end{center}
\end{figure}

To examine how quarantine of symptomatic individuals
affects the reduction in final size of the outbreak relative to the
observed baseline, we assume that
some fraction of symptomatic and infectious individuals, $f_{\rm quar}$, are
moved to completely effective quarantine immediately upon showing symptoms.
We assume that the symptomatic fraction of norovirus cases is
50\%~\cite{sukhrie2012nosocomial,zelner2010infections}.
The results are shown in Figure~\ref{fig:interventions}.

We also examine how environmental cleaning 
affects the outbreak final size, by proportionately scaling both
the environmental transmission rate, $\beta_W$, and the environmental decay rate, $\xi$,
by a factor $\sigma$, where $0\le\sigma\le1$.
The results are shown in Figure~\ref{fig:interventions}.

We also examine the effect of hand washing by proportionately scaling the
direct transmission rates, $B_{ij}$, and the environmental transmission rate, $\beta_W$, by a factor,
$\rho$, where $0\le\rho\le1$.
The results are shown in Figure~\ref{fig:interventions}.

\section{Discussion}

Our estimate of the overall reproduction number for the cruise ship outbreak is
\rtotresult.
There have been few past estimates of the reproduction number of norovirus outbreaks; examination of
norovirus outbreaks in hospitals and long term care facilities have estimated the reproduction number
to be between $1.1$ to $3.4$~\cite{sukhrie2012nosocomial,vanderpas2009mathematical}, while
examination of an outbreak at a large boy scout jamboree estimated the reproduction number
to be ${\cal{R}}_0=7.26$ with 95\% CI $[5.26,9.25]$~\cite{heijne2009enhanced}.  The larger
reproduction number at the jamboree was likely due to the more active mixing of the population,
unlike a hospital or long term care facility where residents are largely bed-ridden.
Our estimated reproduction number is even larger than
that observed at the jamboree, perhaps due to increased probability of
transmission in a population primarily composed of older adults, with potential co-morbidities.

Our modelling analysis finds that only around 8\% and 5\% of norovirus cases in
passengers and crew, respectively, were identified by the ship's infirmary in the outbreak.
Part of the reason may be due to the high asymptomatic rate in norovirus infection;
the overall fraction of norovirus cases that are asymptomatic in outbreaks
has been estimated by past analyses
to be up to 50\% or more~\cite{sukhrie2012nosocomial,zelner2010infections}.
Differing health
status of the crew (on average, being typically younger than passengers)
may also play a role in the disparity seen between the two groups, by
perhaps resulting in a higher symptomatic fraction in the older passengers.
A study of patients and staff in a hospital norovirus outbreak showed that, of the
individuals who tested positive for norovirus infection, fully $23/26=88\%$ of
infected staff were asymptomatic, compared to $4/10=40\%$ of infected patients,
indicating that health status likely does play a role in the rates of asymptomatic
infection~\cite{gallimore2004asymptomatic}.

Beyond the high asymptomatic fraction, however, is the historical evidence for low rates
of cruise passengers with symptomatic gastroenteritis seeking health care aboard the ship;
past survey studies of cruises where outbreaks of norovirus have occurred have
shown that up to 70\% of symptomatic passengers either delayed
reporting to the infirmary, or did not report at all~\cite{neri2008passenger,wikswo2011disease}.
The reasons for the avoidance of health care were in part due to individuals
feeling their symptoms were not serious enough to warrant treatment, and
also due to a desire to avoid enforced quarantine~\cite{wikswo2011disease}.
In addition, crew members on many ships can suffer economic disadvantages if they take time
off due to illness; according to the Glassdoor employer review website ({\tt www.glassdoor.com}, 
accessed April, 2017), where employees can leave anonymous reviews of their employers,
several major cruise lines are reported to be rather less than accommodating when it comes
to paid sick leave for crew members.  This may also help to explain why the fraction of
identified cases is lower in crew than in passengers.

The high asymptomatic fraction, and low rates of health-seeking behaviour, pose problems for
control measures that rely on immediate and complete quarantine of infectious passengers who
have been identified by ship infirmary surveillance as being symptomatic.
Our model predicts that only around 8\% of infectious cases are actually detected by the 
ship's infirmary;
if all of these infectious cases are detected and immediately quarantined, 
our model predicts that this would result
in an equivalent relative reduction in 
the final size of only 0.2\%.  
If we assume an asymptomatic fraction of 50\%, and in a very unrealistic scenario somehow
manage to achieve immediate and complete quarantine of every single newly symptomatic case, the
relative reduction in final size is still only 16\% (see Figure~\ref{fig:interventions}).
Quarantine alone is thus not likely an effective strategy for the control of norovirus transmission.

The 95\% CI on the 
basic reproduction number for environmental transmission was found to be
\renvresultc,
thus a value of ${\cal{R}}_0^{\rm environ}$ greater than 1 is not statistically
excluded by our analysis, and it is likely possible to achieve a sustained outbreak without
direct transmission. Indeed, such sustained transmission due to fomite contact alone
has been noted among consecutive groups of different people renting a houseboat, where there was
no common population or direct contact between the consecutive groups~\cite{jones2007role}.
Environmental cleaning thus has a role to play in control of sustained outbreaks.
However, 
it is worth noting that the cruise ship on which the outbreak occurred, like all cruise
ships, had to pass a rigorous sanitation inspection~\cite{lawrence2004outbreaks}, and thus it was already a
quite clean environment.  
It only takes a few norovirus particles to transmit infection~\cite{dolin2007noroviruses},
and the pathogen is notoriously difficult to kill on surfaces~\cite{feliciano2012efficacies}. 
It has been shown that
wiping surfaces with common detergents served more
to simply spread the pathogen around, rather than killing it~\cite{barker2004effects}.
Additionally, we find that 
direct transmission appears to be the overwhelmingly dominant factor in outbreak size,
and we find that even the most stringent cleaning that
eliminated all the virus from the environment would result in a relative reduction in outbreak size
over the two cruises of only $3.5$\%.

Both the direct and environmental transmission can be
reduced through personal hygiene measures, such as
hand washing, that reduce the probability
of transmission upon contact (thus proportionally reducing both ${\cal{R}}_0^{\rm direct}$ and
${\cal{R}}_0^{\rm environ}$).
Our analysis thus indicates that aggressive educational campaigns aimed at improved
hand washing practices would likely be most efficacious in reducing the morbidity burden, and effective, widespread hand washing can entirely prevent a potential outbreak.
Indeed, the analysis of a norovirus outbreak at a boy scout jamboree found that the
implementation of rigorous hand washing protocols reduced the reproduction number by 85\%~\cite{heijne2009enhanced}.
A survey analysis of a past norovirus outbreak aboard a cruise ship
also found that lackadaisical attitudes towards hand hygiene was one of the dominant risk factors affecting
probability of infection during the outbreak~\cite{neri2008passenger}.

However, unlike children at a jamboree, who can be forced by adult authorities
to wash their hands before eating, overcoming
long-standing poor hygiene habits among some adults on a cruise ship can be a challenge~\cite{curtis2009planned}.
Informational signs in bathrooms, and at entrances to ship eating areas, promoting the
importance of proper hand washing might perhaps be useful, particularly if the signs stress the probable loss
of quality vacation time for passengers who fall ill, while also pointing out that studies
have shown that passengers who don't wash their hands before eating are much more likely to fall ill.

\section{Summary}

We have presented a novel mathematical model for norovirus disease transmission aboard a cruise ship that
includes both direct and environmental transmission.  This is not the first
model for a disease to include environmental transmission; for example, some cholera models
include both direct and environmental transmission~\cite{tien2010multiple,tuite2011cholera}.
However, to the authors' knowledge, this is the first time that such a model has been used to
quantify the relative contribution of direct and environmental transmission for norovirus disease.
With the quantification of these relative contributions to transmission, 
the relative efficacy of potential control strategies aimed at either
environmental sanitation, personal hygiene, or quarantine can be assessed.

We find that due to the high asymptomatic fraction of norovirus infection, and
low rates of health-seeking behaviour,  quarantine of symptomatic passengers
aboard a cruise ship is likely ineffectual at outbreak control.
We also find that the rates of environmental transmission are high enough to likely result in sustained
outbreaks, but that overall transmission is dominated by direct person-to-person contact.
Thus, environmental cleaning is likely to have little impact on the final size of outbreaks
on cruise ships.  These findings are supported by past qualitative observations that
norovirus outbreaks aboard cruise ships are notoriously difficult to control~\cite{hadjichristodoulou2011surveillance,bert2014norovirus,wikswo2011disease,lawrence2004outbreaks}. 

We find that reduction in environmental and direct transmission is best achieved
with personal hygiene measures, such as rigorous hand washing, designed to reduce the probability
of infection upon contact with a contaminated surface or infectious individual.


\clearpage
\appendix
\section*{Appendices}
\renewcommand{\thesubsection}{\Alph{subsection}}


\end{document}